# A Longitudinal Analysis on Instagram Characteristics of Olympic Champions

Amirhosein Bodaghi

Federal University of Rio de Janeiro, Department of Computing Science, Centre of Mathematical and Natural Sciences – CCMN, Rio de Janeiro, Brazil
bodaghi@ppgi.ufrj.br

## Abstract

This study examines Olympic champions' characteristics on Instagram to first understand how age and gender affect the characteristics and their interrelations and second to see if the future changes in those characteristics are predictable. We crawled Instagram data of individual gold medalists in Rio2016 Olympic for a period of 4 months and utilized a content analytic method to analyze their photograph posts. The cross-sectional analysis shows as the champions get older, the rate of follower engagement decreases in both genders, but men increase their pure self-presentation posts while women extend their circle of followings. In its approval, the longitudinal analysis shows when the higher rate of engagement is achieved, then men and women champions lose their tendency in increasing self-presenting posts and number of followings respectively. These findings in the light of relative theories and the previous literature contribute to a better understanding of athletes' cyber behavior in social media. Moreover, the findings serve as a guide for sport researchers seeking to grasp the ways that aid athletes to better interact with their followers and build the personal brand, which involves sponsorship and other promotional opportunities.

# 1. Introduction

Among the variety of benefits have come to the athletes by the new technologies which run the gamut from health aids (Bodaghi, 2016) to marketing (Park et al., 2020), those provided by social media are undeniable. Social media as the flagship of new computer-based technologies draw athletes since it provides a way for them to expand their influence, reputation and earning power (Farrington et al., 2014). Many athletes use social media platforms to build their own individual brands in order to market themselves to potential sponsors or corporate partners (Mackova and Turkova, 2019). Karg and Lock (2014) came to this conclusion that while social media may not directly yield revenue, by establishing fan communities they confer the advantage of brand awareness to sport entities. Especially, it is of great importance for well-known athletes who do not receive much mainstream media coverage to take advantage of social media to generate publicity, as this is often their only way to make their personal brand (Geurin-Eagleman, 2013).

As social media evolves into more visual content, interactive engagement and discussion can be driven by sharing of photos and videos (Marshall, 2010). The social media bring freedom for individuals to choose content and come up with different identities (Bullingham and Vasconcelos, 2013). These cyber-identities impact all the benefits and harms that use of social media brings to athletes. Hence, the cognition of these cyber-identities is a fundamental step to the realm of research about athletes on social media. In this vein, the relationships between cyber characteristics by which the athletes have crafted their virtual self over time, are of great importance. This study tapped into this topic by conducting a longitudinal research on Instagram accounts of Rio Olympic gold medalists to find possible relationships between different characteristics of these champions. The Rio Olympics is one of the first mega events where there has been full application of the social media and the possible issues and trends that will influence these types of events in the future (Lee et al., 2019).

Therefore, the purpose of this study is to examine Olympic champions' characteristics on Instagram to first understand whether differences exist between them with respect to gender and age, and then to track the change in these characteristics over time with the hope of predicting future changes in these characteristics based on the current data. Hence, this study is intended to answer the following research questions:

RQ1: How do age and gender affect the champions' characteristics and their interrelationships?

RQ2: Are the future changes in the champions' characteristics predictable by their current values or rate of change?

To answer the above questions we have gathered data of Olympic gold medalists out of their Instagram accounts (where available) in different observations over time, and then conducted two types of analyses. First, a cross-sectional analysis in which we explored data of each observation separately but then merged the results to get insights particularly to answer RQ1. Second, a longitudinal analysis to gauge the rate of changes over time that let us examine possible correlations between future and past as demanded by RQ2.

## *1.1. Theoretical and Practical Implications*

This study opens up new doors in the search for potential applications of well-known theories such as evolutionary psychology and natural selection in the realm of cyber behaviors, particularly by considering natural features like gender and age. The findings also pave the way for the new extensions of the theory of uses and gratification in social media with focusing on celebrities at the level of Olympic gold medalists.

From the practical point of view, the ability to know the interrelations between user characteristics and predict their future change aids to detect possible correlations between different cyber behaviours. For instance, higher rates of self-presentations as a sign of narcissism may appear with marked change in the rate of mutual friendship as a symptom of low-level self-esteem. When it comes to the Olympic champions, the importance of these findings gets bold. For example, what change in user characteristics would yield higher rates of engagement from the followers to boost the athletes brand or how to pre recognize the signs of depression or low self-esteem for a professional athlete before the onset of great contests such as the Olympic.

Of course, the establishment of any link between a certain behavior/trait and a specific characteristic demands exclusive research and studies like what we have witnessed in the literature for the case of narcissism and self-presentation, however, it may not render a standard measure for all genders, ages, and professions. Only by looking through the lens of these different features and narrowing the target (as this study focuses only on Olympic gold medalists), we can see the balance among the values of users' characteristics from which a certain trait may rise up. Therefore, studying the interrelations between user characteristics particularly in a dynamic way with considering the impact of time is a priory for further research in this regard.

## 2. Literature Review

Based on the context of this paper we have divided the literature into three subsections; engagement, age and gender, and self-presentation, which form the base of this, study with respect to the social media characteristics.

## *2.1. Engagement*

Visual-based social media are growing exponentially and have become an integrated part of the customer engagement strategy of many brands. The action of being engaged is a general concept but when it comes to social media, it mostly takes the forms of liking or commenting that indicate some sort of involvement with the posts. Schaffer et al. (2019) assessed Instagram use across cultures and found that the global cross-cultural differences may impact the reasons why individuals use Instagram. Arman and Pahrul Sidik (2019) studied the rate of engagement in Instagram by focusing on Indonesian government ministry and institutions and found that the higher number of followers leads to the lower rate of engagement. Oliveira & Goussevskaia (2020) analyzed sponsored content and its impact on user engagement in the context of influencer marketing on Instagram. They found that an increasing amount of sponsored content could negatively affect audience engagement with an influencer's profile.

Ferrara and Yang (2015) focused on online popularity and topical interests through Instagram. Their analysis provided clues to understand the mechanisms of users who interact in online environments and the way collective trends emerge from individuals' topical interests. Gaspar and Tietzmann (2018) examined the ways in which the number of followers in the official profiles of Brazilian Olympic athletes in the social networks Facebook, Twitter and Instagram was impacted by their results and their participation in the Rio 2016 Games. They found the performance of the athletes in the last Olympic Games influenced the growth of the number of followers in their social networks, with Instagram being the network that featured the biggest increase, and gold medalists had the highest growth rates. Rietveld et al. (2020) extract emotional and informative appeals from Instagram posts and find support for the influence of emotional and informative appeals.

Abuin-Penas and Martinez-Patino (2019) studied the athlete communication on Instagram by focusing on the communication strategies of the Spanish Winter Olympic team during 2018, when the PyeongChang Olympic Games were held. They noticed Instagram is revealed as an effective tool for Olympic athletes when it comes to generating visibility among their audiences and fan engagement increases during the celebration of an international mega event as the Olympic Winter Games. Romney and Johnson (2020) study sought to understand what types of images generate more engagement from social media audiences. They performed a content analysis of nearly 2000 images shared by Sports Networks on Instagram and found that images that contained narrative or Meta communicative messages resulted in greater interest and engagement by audiences through the manifestation of likes and comments.

## *2.2. Gender and Age*

Butkowski et al. (2020) conducted a content analysis to investigate gender display, as rituals of gender behaviour, in young women's Instagram selfies alongside its relationship to feedback, such as likes and comments. They found women who incorporate and exaggerate gender displays in their selfies tend to receive more feedback. Therefore, they suggested that gender stereotyping in Instagram selfies is related to reinforcing feedback. Ferwerda and Tkalcic (2018), looked at personality prediction from Instagram picture features and show that visual (e.g., hue, valence, saturation) and content features (i.e. the content of the pictures) can be used to predict personality from and perform in general equally well. Han et al. (2018) conducted a comprehensive analysis on social media use and engagement by age and gender on Instagram. They defined five user age groups (from 10s to 50s) and two user gender groups (males and females), and compared them based on three aspects: activity, image object, and tag. Their study results indicated that each user group exhibits unique characteristics and the features from each aspect can be used to develop user classification models (82% for gender and 43% for age classification) without relying on the information that specifically indicates age and gender.

Jang et al. (2015) showed the possibility of detecting age information in user profiles by using a combination of textual and facial recognition methods. Song et al. (2018) presented a study of two personal characteristics - age and gender - related to user engagement on Instagram that can be determined through the characterization of images and tags. They demonstrated the strong influence of age and gender on Instagram use in terms of topical and content differences.

### *2.3. Self-presentation*

Having regard to the sociotechnical aspects of using social networks, a series of research has been particularly done on Instagram. Sheldon and Bryant (2016) investigated motives for using Instagram and its relationship to contextual age and narcissism. Jang et al. (2015) analyzed 'like' activities in Instagram. They provided an analysis of the structural, influential and contextual aspects of like activities from the test datasets of users on Instagram. Bij de Vaate et al. (2018) aimed to profile selfie-makers' motivations and behavior, and examine the extent to which underlying mechanisms preceding selfie-posting are interconnected. They found entertainment and moment-retention as main motivations for selfie-making.

A series of research has been focused on athlete self-presentation on social media platforms, and results revealed that athletes engage in backstage performances, people engage in these behaviors when no audience is present, to discuss their private lives and engage with individuals (Burch et al., 2014). Xu and Armstrong (2019), explored the gendered differences between the self-portrayals of U.S. and Chinese athletes at the 2016 Rio Olympics. Their results suggested that cultural background had a substantial impact on self-representation for all participants. Devonport et al. (2019) by adopting a feminist perspective explored two research questions: (1) how do male and female athletes perform an athletic identity through photographic self-representation, and (2) what are the messages they look to convey, as role models, through these images? They found Participants typically wanted to appear in action shots, emphasizing good technique, displaying a sporting physique and in relevant uniforms. Findings suggested that whilst athletes sought to champion their sport and the physical and psychological qualities that participation produces, gendered performances were also evident in production and interpretation of many images, thus highlighting the pervasive nature of gendered sporting participation.

Geurin-Eagleman and Burch (2016) presented a gender-based analysis of Olympic athletes' self-presentation on Instagram in order to develop an understanding of the ways in which athletes use this medium as a communication and marketing tool to build their personal brand. Szabo and Buta (2019) performed an exploratory work to assess the nature and origin of sport-selfies. After a content analysis of 930 random sport-selfies, they found that while men dominate sports, women post more sport-selfies than men do, with Outdoor and land-based sport activities the most frequent. It also emerged that in contrast to research on selfies in general, sport-selfies are primarily posted by white individuals and most sport-selfies originate from Germany, Brazil, and Russia and images linked to body conditioning are the most popular. Therefore, they concluded that sport-selfies are culture, gender, and sport specific.

Despite all the research that has been centered around the presence of athletes on social media, up to this date no sound research has been conducted to investigate possible relationships between athletes' characteristics on Instagram, within themes of gender and age, and particularly with considering the passage of time. However, our findings, which are based on a longitudinal study, open new insights into it that would pave the way to advance further investigations about the athletes' cyber characters.

## 3. Method

We utilized a content analytic method to analyze Olympic gold medalists' photographs on Instagram. As indicated by Riffe, et al. (2005), content analysis is an efficient and replicable method for examining content, both written and visual. The Olympic of Rio 2016 was held for 16 days from 5 to 21 August. There were 42 summer sports with 306 events. The number of gold medals were given to the individual champions (excluding team sports) were 226. And since some athletes were champion in more than one event, the total number of individual gold medalists was 213 (https://www.olympic.org), from which 144 (83 men and 61 women) gold medalists had a publicly available Instagram account in all of the observations during the 4 months period of data gathering (Table 1).

**Table 1.** Individual Events and Gold Medalists of Olympic Rio2016.

| Sports With Individual Events | Men Individual Events | Women Individual Events | Gold Winner (Men) | Gold Winner (Women) | Gold Winners with active Instagram Account | |
|---|---|---|---|---|---|---|
| | | | | | Men | Women |
| Archery | 1 | 1 | 1 | 1 | 0 | 0 |
| Athletics | 22 | 21 | 20 | 20 | 16 | 14 |
| Badminton | 1 | 1 | 1 | 1 | 0 | 1 |
| Boxing | 10 | 3 | 10 | 3 | 8 | 2 |
| Canoeing | 6 | 3 | 6 | 3 | 4 | 1 |
| Cycling | 7 | 7 | 6 | 7 | 5 | 7 |
| Diving | 2 | 2 | 2 | 2 | 0 | 0 |
| Equestrian | 2 | 1 | 2 | 1 | 0 | 1 |
| Fencing | 3 | 3 | 3 | 3 | 2 | 2 |
| Golf | 1 | 1 | 1 | 1 | 1 | 0 |
| Gymnastics | 8 | 7 | 7 | 5 | 4 | 5 |
| Judo | 7 | 7 | 7 | 7 | 6 | 4 |
| Pentathlon | 1 | 1 | 1 | 1 | 1 | 1 |
| Rowing | 1 | 1 | 1 | 1 | 1 | 0 |
| Sailing | 3 | 2 | 3 | 2 | 3 | 2 |
| Shooting | 9 | 6 | 8 | 6 | 2 | 2 |
| Swimming | 14 | 14 | 12 | 11 | 12 | 8 |
| Table Tennis | 1 | 1 | 1 | 1 | 0 | 1 |
| Taekwondo | 4 | 4 | 4 | 4 | 2 | 2 |
| Tennis | 1 | 1 | 1 | 1 | 1 | 1 |
| Triathlon | 1 | 1 | 1 | 1 | 1 | 1 |
| Weightlifting | 8 | 7 | 8 | 7 | 6 | 2 |
| Wrestling | 12 | 6 | 12 | 6 | 8 | 4 |
| Sum | 125 | 101 | 118 | 95 | 83 | 61 |

In this study, we examined just individual gold medalists because team sports athletes who won gold medals often take advantage of these teams' endeavors and the attention received by the teams/leagues. While individual gold medalists often rely on their own efforts to establish their personal brand and to remain relevant in the intervals between Olympic Games. Then for each user, we gathered the following information: 1-the number of posts; 2-the number of followers; 3- the number of followings; 4-the max number of likes and comments among the last ten posts; 5- the number of self-presenting posts (posts in which the champion face is detectable) from the previous ten posts; 6- the number of pure self-presenting posts (posts in which the champion stands alone) from the previous ten posts; 7- gender; 8- age.

This research only considered photo posts and if there were any videos or multiple-pictures (as one post) in the last ten posts then we went back to more previous posts till we found ten photo posts. Another point to mention is the notion of "selfie" which despite its popularity in Instagram, it is the least popular photograph athletes include on their feed (Smith and Sanderson 2015). Thus, to overcome the constraints of selfie, which solely focus on the subject of the photograph, we defined self-presenting posts as all photos in which the athletes' face is recognizable, while for counting pure self-presenting posts, the condition of being solitary in the photo was added. In addition, the rate of engagement is defined as the sum of the number of likes and comments achieved by the last photo posts.

All the above information gathered for the champions distilled into 4 characteristics that forms the foundation of this study. Since this rate of engagement and the number of followings both are dependent on the number of followers, we normalize them by dividing to the number of followers and consider these proportions i.e. fw/fr and eng/fr as two characteristics of this study. The other two characteristics are sp and psp/sp which both are in the form of ratio, as the former indicates number of self-presenting posts among the last 10 photo posts and the later refers to the ratio of pure self-presenting posts among those self-presenting posts.

### *3.1. Data and Codes*

The process of gathering information by the introduced method took four rounds consecutively in August, September, October and November 2019 and each lasted 4 days from 9 to 12 of the months. All done by human referees with meticulous examining and exploiting data from each Instagram account. The dataset is publicly available on the Internet. In addition, the codes for data analysis, which we will get to in the next section, all, have been written by python and are publicly available on Github.

## 4. Results

We have done this study at two stages: in the first stage, we performed a cross-sectional analysis that seeks to investigate relationships between the characteristics at each observation separately along with some explorations with regard to the demographic data such as gender, age, country and type of sport. In the second stage, we run a longitudinal analysis to see how the athletes' characteristics change over time and how these changes are related to each other and the prior values of the characteristics at the beginning of each time interval.

## *4.1. Cross Sectional Analysis*

As mentioned above, in the cross sectional analysis we look at the observations as independent snapshots of data and study the characteristics regardless of any change over time. This cross sectional study divides to 2 subcategories of analysis as correlations, gender, age, and country and sport.

### *4.1.1. Correlations*

At first, we examine the possible correlations between the characteristics at each observation. Fig. 1 shows the results of Pearson and Spearman correlation between the characteristics (in addition to the age of athletes) for each observation.

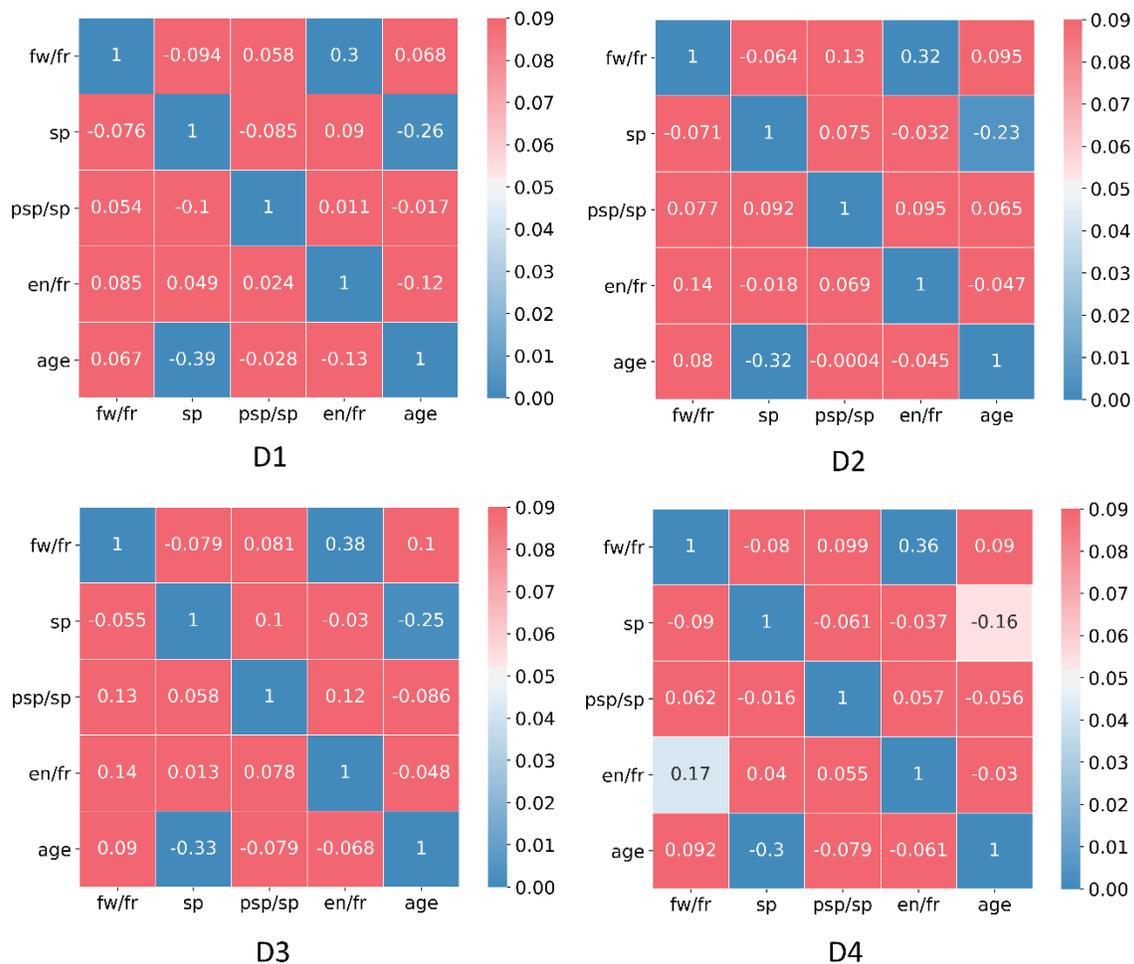

**Fig. 1** Correlation between user characteristics at each observation. D1, D2, D3 and D4 denote the 4 observations of the study. For each observation, the lower triangle of the square presents the values of Pearson correlation and the upper triangle belongs to the values of Spearman correlation. The colour of each cell refers to the level of P-value for the correspondent correlation. For example in D4 the Pearson correlation between age and sp is -0.3 with a p-value less than 0.02, which proves the significance of that relationship.

In looking at the Fig. 1, there appears to be two consistent patterns in all of the four observations. First, between age and sp that for both Pearson and Spearman correlations we see negative and of course significant coefficients that refer to the existence of an inverse monotonic and linear relationship between age and self-presentation. In other words, the older the champion is the less self-presenting photos she/he would post on Instagram. The second consistent pattern is between fw/fr and en/fr with a positive and significant spearman correlation, which denotes the existence of a positive monotonic relationship between them. This means a champion with more reciprocal ties with his/her followers would draw more engagement from them. In addition, an interesting point obvious from these observations is that there is no relationship between sp and psp/sp that signals the importance of psp/sp as a new characteristic or at least its independence from the sp as a measure of self-presentation.

*4.1.2. Demographic Study*

Gender, age, nationality and professionals (type of sport) are among the most important demographic features and it is of interest to see how the characteristics of Olympic champions differ based on them. Since the 4-month period of the study does not change the age of champions by even a single unit from the first observation, and gender, nationality and type of sports are constant; we merge the 4 observations and study the whole at once. In other words, instead of exploring 4 matrices of size 144×4, where 144 users (rows) have their characteristics' values in 4 columns, we study the matrix of 576×4 to see how the demographic features affect the characteristics of champions on Instagram.

*4.1.3. Gender*

Fig. 2 shows the results for gender analysis by KDE (Kernel Density Estimate) plots of paired characteristics along with the distribution of each characteristic. The plots are drawn by the Seaborn module of Python.

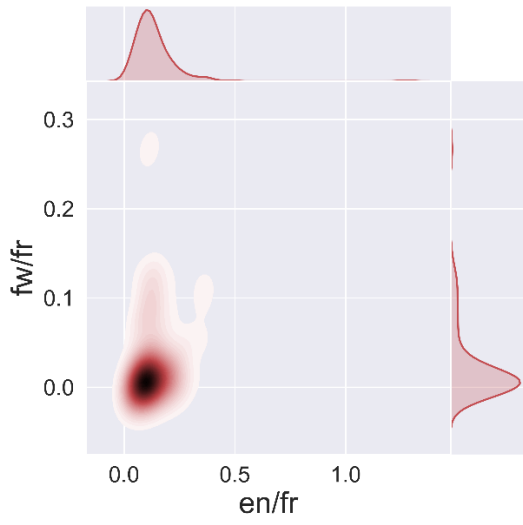
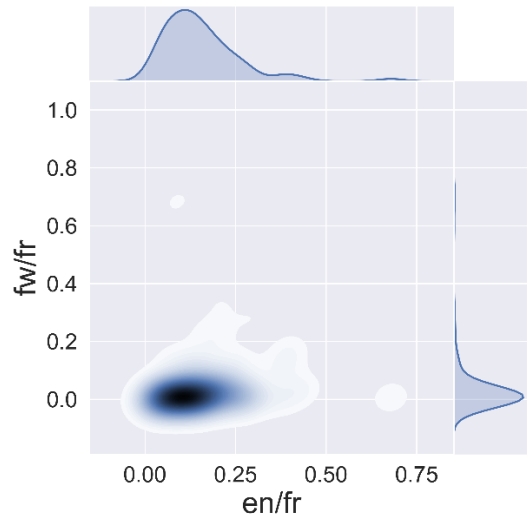

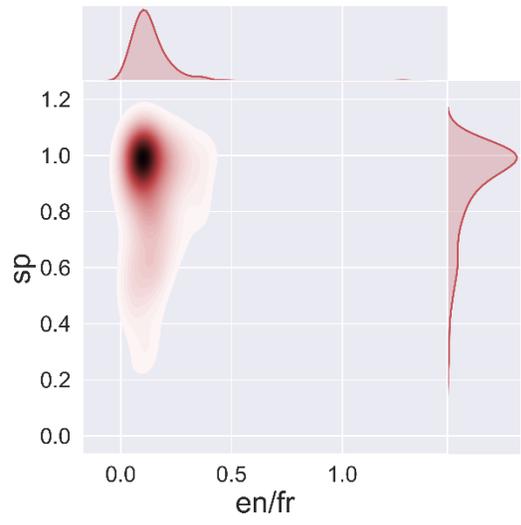
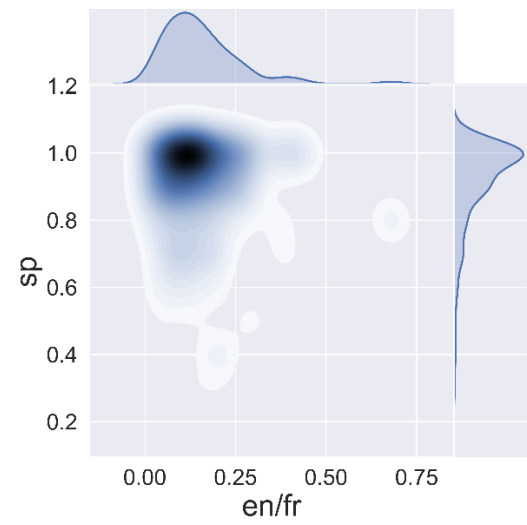

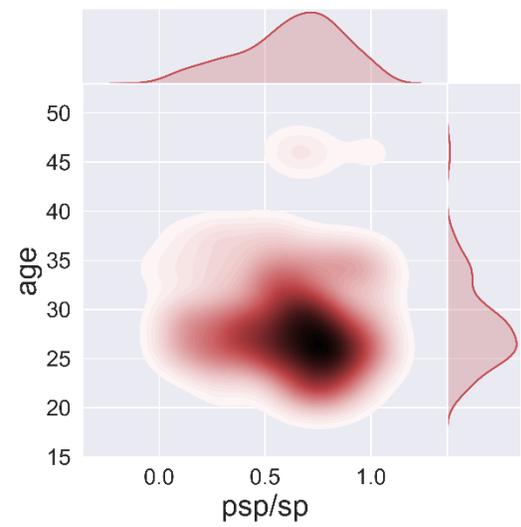
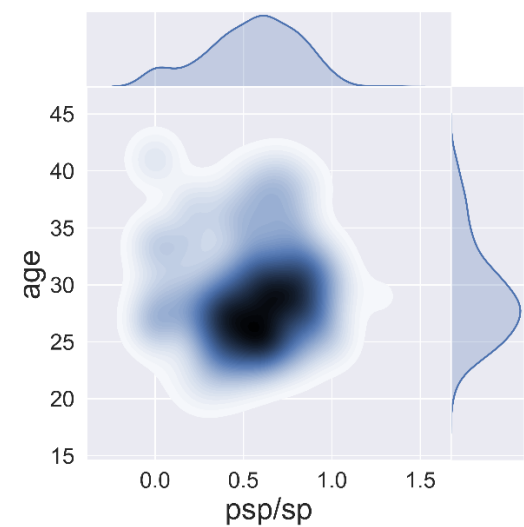

**Fig. 2** Gender-based KDE plots of paired characteristics.
The red and blue plots respectively belong to women and men.

Even though, in general, men and women champions have many similarities in terms of their characteristics but by detailed looking at Fig. 2 some subtle differences would appear. For men, the characteristics of en/fr and fw/fr seem to be stretched in a wider range than for women. A fact that en/fr and fw/fr go hand in hand is quite in line with the previous finding about the correlation between them, but having this effect more stressed in men is a subject of discussion. With respect to the characteristic of sp, women come up with a wider range than men do and the point is that this stretch is toward lower values of sp. In other words, the chance of seeing a woman champion with low levels of self-presentation is higher relative to her male counterparts. Besides this, psp/sp, which is more stretched for champions younger than 30 years old, appears to have higher frequencies for women than men. In effect, women champions older than 30 years old, show a tendency to take their self-presentation photos with others and not necessarily stand alone in their post (pure self-presentation).

*4.1.4. Age*

For the age analysis, first we define three ranges as follows:

➢ *Range_1: age < 26*
➢ *Range_2: 26 ≤ age < 30*
➢ *Range_3: 30 ≤ age*

By the above separation, range 1 to 3 respectively include 32, 61, and 51 users that fits to a normal Gaussian distribution. Fig 3 shows the violin plot of the 4 characteristics for each range both for men and women.

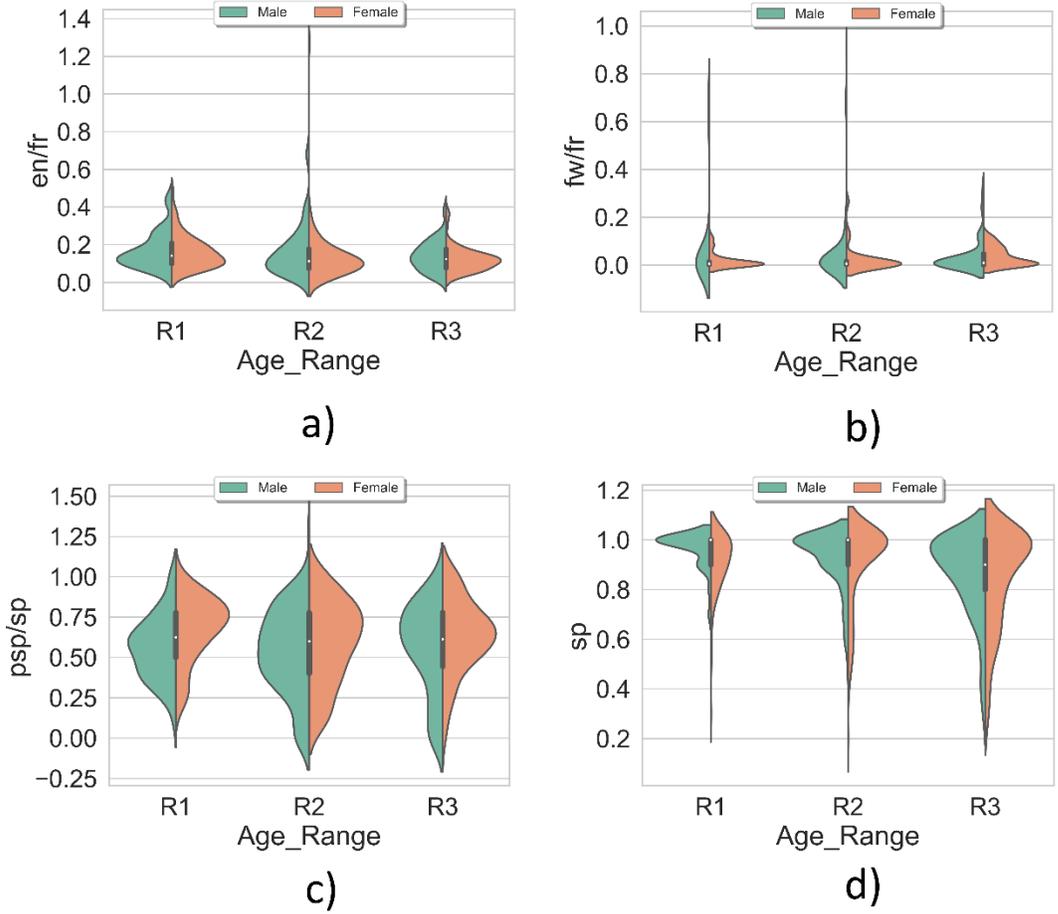

**Fig. 3** Age-based violin plots of the user characteristics.
R1, R2, and R3 respectively refer to the defined age-ranges 1 to 3.

By looking at Fig 3-a we see en/fr for both men and women loses its extension toward higher values and shrinks around lower values (less than 0.2) as the champions gets older, however the position of the distribution's peak is not affected much. It is worth clarifying that the long tails of some distributions, which are due to the outliers demand specific studies and would not be taken into account in the current interpretations. In Fig 3-b two contradictory behaviors have come to surface from men and women. While men experience a contraction in their fw/fr range, as they get older, women stretch it to the higher values. Indeed, aging makes the women champions to strengthen their links to the followers with reciprocation or even laying out new connections to non-followers. In Fig 3-c men and women champions again show two opposite behaviors as they get older, this time in terms of the position of the distribution's peak. While men show an increase in their psp/sp as they get older, women become more reluctant to stand alone in their self-presenting posts as they age and decrease their psp/sp. However, by looking at Fig 3-d we see both men and women champions feel less motivated to post self-presenting posts as they get older and their distributions of sp stretch toward the lower values.

### *4.1.5. Country and Sport*

In this section, we explore data to find the countries and sports whose champions achieved the highest and lowest of the characteristics' values. Table 2 demonstrates the three countries and sports, which gained the highest and lowest ranks in each of the characteristics.

**Table 2.** The countries and sports whose champions achieved the highest and lowest ranks for each of the characteristics.

| Characteristics | Countries (out of 45 countries in total) | | | | | | | | | | | |
|---|---|---|---|---|---|---|---|---|---|---|---|---|
| | Highest Ranks | | | | | | Lowest Ranks | | | | | |
| | 1st | | 2nd | | 3rd | | 1st | | 2nd | | 3rd | |
| | Mean | Var | Mean | Var | Mean | Var | Mean | Var | Mean | Var | Mean | Var |
| fw/fr | Tajikistan | | Kenya | | South Korea | | Colombia | | Puerto Rico | | Armenia | |
| | 0.3044 | 9.08e-4 | 0.1738 | 0.0819 | 0.1399 | 0.0099 | 5.93e-4 | 1.59e-7 | 6.13e-4 | 6.62e-9 | 6.56e-4 | 1.29e-10 |
| sp | Bahamas | | Canada | | Azerbaijan | | Tajikistan | | New Zealand | | Iran | |
| | 1.0 | 0.0 | 1.0 | 0.0 | 1.0 | 0.0 | 0.45 | 3.33e-3 | 0.6125 | 4.1e-3 | 0.6499 | 3.33e-3 |
| psp/sp | Denmark | | Turkey | | Puerto Rico | | Cuba | | New Zealand | | Uzbekistan | |
| | 0.8416 | 0.0038 | 0.8333 | 0.0914 | 0.8111 | 0.0192 | 0.1888 | 0.0127 | 0.3125 | 0.1145 | 0.3710 | 0.0242 |
| en/fr | Armenia | | Bahamas | | China | | Puerto Rico | | South Africa | | Uzbekistan | |
| | 0.3699 | 2.95e-4 | 0.3327 | 0.0168 | 0.3097 | 0.0746 | 0.0324 | 1.34e-5 | 0.0518 | 2.65e-4 | 0.0521 | 3.9e-4 |
| | Sports (out of 109 sport-event in total) | | | | | | | | | | | |
| Characteristics | Highest Ranks | | | | | | Lowest Ranks | | | | | |
| | 1st | | 2nd | | 3rd | | 1st | | 2nd | | 3rd | |
| | Mean | Var | Mean | Var | Mean | Var | Mean | Var | Mean | Var | Mean | Var |
| fw/fr | Boxing Heavyweight | | 3000 metres steeplechase | | Table tennis | | Gymnastics Individual all-around | | Weightlifting 77 kg | | Swimming 200 m butterfly | |
| | 0.7370 | 0.0137 | 0.6492 | 0.0063 | 0.2665 | 1.15e-6 | 2.79e-5 | 4.83e-13 | 4.94e-5 | 2.10e-12 | 6.12e-5 | 2.16e-13 |
| sp | Judo Lightweight | | Fencing Individual sabre | | Canoeing Sprint C-1 1000 m | | Shooting 50 metre pistol | | Wrestling freestyle 63 kg | | Hammer throw | |
| | 1.0 | 0.0 | 1.0 | 0.0 | 1.0 | 0.0 | 0.35 | 0.0033 | 0.525 | 0.0291 | 0.5625 | 0.0169 |
| psp/sp | Long jump | | Canoeing Slalom K-1 | | Heptathlon | | Rowing Single sculls | | Swimming 200 m butterfly | | Fencing Individual foil | |
| | 0.9736 | 0.0023 | 0.8944 | 4.11e-5 | 0.8916 | 0.0082 | 0.0 | 0.0 | 0.0357 | 0.0051 | 0.05 | 0.01 |
| en/fr | Swimming 200 m breaststroke | | Modern pentathlon | | Swimming 200 m freestyle | | Golf | | Boxing Light welterweight | | Weightlifting 62 kg | |
| | 0.4346 | 1.54e-5 | 0.3920 | 0.1442 | 0.3827 | 0.1007 | 0.0183 | 2.54e-5 | 0.0287 | 2.97e-6 | 0.0288 | 1.20e-7 |

The findings presented in this table give only a flavor of what disparities exist between the champions' characteristics in terms of profession and nationality but should be studied through the lens of cultural and professional contexts to yield fruitful interpretations.

### *4.2.   Longitudinal Analysis*

In the longitudinal analysis, we focus on the change that users' characteristics experience over time. Prior to the longitudinal analysis, the characteristics' mean and variance for each observation were gauged with respect to the genders to keep track of their progress. The results are shown in Fig. 4.

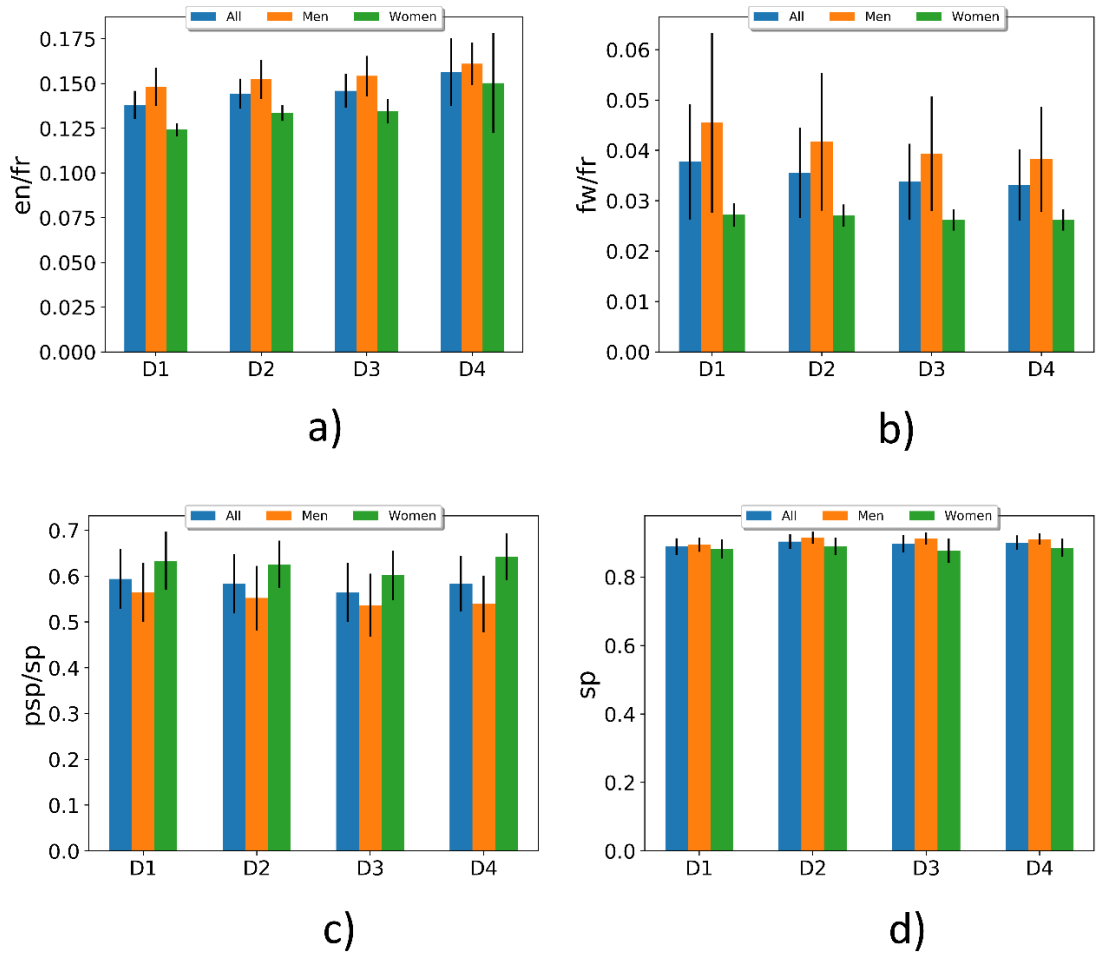

**Fig. 4** The mean and variance of user characteristics for different observations

The obvious pattern in Fig. 4 is the smooth increase of en/fr for men champions, which is accompanied by a smooth decrease in fw/fr over the course of these 4 observations. Although, such increase in en/fr is recognizable for women champions also, but that does not go along with a tangible change in fw/fr. For the other two characteristics, no consistent pattern can be seen over the observations.

In our longitudinal analysis over the observations, we have two objects to seek out. First, the relationship between the amount of change champions' characteristics experience in their values over time. Second, the relationship between these changes over time with the prior values of the characteristics. The possible findings out of these two objects would give us insights into the future of champions' Instagram activities. Since we have 4 observations in this study, there would be three intervals to calculate change over them. The intervals '21', '32', and '43' respectively refer to intervals between observations '1 and 2', '2 and 3', and '3 and 4'. The matrices for each observation ($D_m$) and each interval ($D_{m(m-1)}$) are achieved by the following equation.

$$D_m = \begin{bmatrix} a_{1,1}^m & a_{1,2}^m & a_{1,3}^m & a_{1,4}^m \\ \vdots & \vdots & \vdots & \vdots \\ a_{144,1}^m & a_{144,2}^m & a_{144,3}^m & a_{144,4}^m \end{bmatrix} \quad, \quad D_{m(m-1)} = \begin{bmatrix} b_{ij} \end{bmatrix} \quad where \quad b_{ij} = a_{ij}^m - a_{ij}^{m-1} \quad (1)$$

$m$ : the number of observations $\rightarrow m \in [1, 4]$

144 ($rows$) : $number\ of\ users$, $and$ 4 ($columns$) : $number\ of\ characteristics$

The results of correlations over the longitudinal data are presented in Table 3. All correlations are calculated by Pandas library of Python and the function of *pandas.DataFrame.corr* that computes pairwise correlation of columns. The upper half of the table shows the correlation coefficients of rate of change for a characteristic with others between two consecutive intervals (i.e. '$D_{21}$ vs. $D_{32}$' or '$D_{32}$ vs. $D_{43}$'). The lower half of the table does the same but between an interval and the primary values of that interval (i.e. '$D_1$ vs. $D_{21}$', '$D_2$ vs. $D_{32}$', or '$D_3$ vs. $D_{43}$'), which are the beginning observations for that interval.

Furthermore, the table only includes those pair of characteristics whose correlations' p-values, either for Pearson or Spearman, have been less than 0.05 in at least one column, for all the pairs of interval-interval in the upper half of the table (i.e. 21-32, and 32-43) or for at least two of the pairs of observation-interval in the lower half of the table (i.e. 1-21, 2-32, and 3-43). For instance, the pair of psp/sp and fw/fr has come to the table only because in the column of Pearson correlation in Age-R2 for both pairs of interval-interval (i.e. 21-32 and 32-43) the P-values of the correlations had been less than 0.05. In other words, for Age-R2 users, the significance of an inverse linear relationship between the rates of change in psp/sp in one interval with the rates of change in fw/fr in the next interval is proved regardless of the number of intervals. In another example, we see the pair of fw/fr and en/fr in the table merely due to the significance of Pearson correlation for women in two cases (if it was not for at least two cases we would not have it in the table): 1- between the first observation ($D_1$) and its subsequent interval ($D_{21}$), 2- between the third observation ($D_3$) and its subsequent interval ($D_{43}$). This signals the existence of a linear relationship between the values of fw/fr for women champions with the amount of change in en/fr they are about to experience in the following month.

**Table 3.** Correlations over the longitudinal data. The upper half of the table presents correlation values between rates of change in the characteristics over the observations' intervals for champions of different gender and age. The lower half of the table focuses on the correlations between rates of change in the intervals with the primary values in the beginning of the intervals. Merely those correlation values (coefficients) are added to the table whose p-values were less than 0.05. For the sake of simplicity, the terms of 'observation' and 'interval' are omitted from the table, so for instance 1-21 means correlation between observation 1 and interval 21.

| | | | Gender Analysis | | | | Age Analysis | | | | | |
|---|---|---|---|---|---|---|---|---|---|---|---|---|
| | | | Men (83 users) | | Women (61 users) | | Age_R1 (32 users) | | Age_R2 (61 users) | | Age_R3 (51 users) | |
| | | | Pearson | Spearman | Pearson | Spearman | Pearson | Spearman | Pearson | Spearman | Pearson | Spearman |
| Rate of Change vs. Rate of Change | en/fr & sp | 21-32 | -0.2411 | | | | | | -0.3064 | | | |
| | | 32-43 | -0.4788 | | | | | | -0.2960 | | | |
| | psp/sp & psp/sp | 21-32 | -0.2530 | -0.2075 | | | | | | | | |
| | | 32-43 | -0.5481 | -0.2136 | | | | | | | | |
| | en/fr & en/fr | 21-32 | -0.4290 | | | | | | | | | |
| | | 32-43 | -0.2398 | | | | | | | | | |
| | fw/fr & fw/fr | 21-32 | | 0.3281 | 0.2735 | 0.3933 | | 0.4813 | | 0.3456 | | 0.2875 |
| | | 32-43 | | 0.4180 | 0.6110 | 0.3315 | | 0.4530 | | 0.4785 | | 0.3099 |
| | sp & sp | 21-32 | | -0.2081 | | | | | | | | |
| | | 32-43 | | -0.2924 | | | | | | | | |
| | psp/sp & sp | 21-32 | | | | | | -0.3515 | | | | |
| | | 32-43 | | | | | | 0.4727 | | | | |
| | psp/sp & fw/fr | 21-32 | | | | | | | -0.4307 | | | |
| | | 32-43 | | | | | | | -0.2445 | | | |
| Rate of Change vs. Early Values | psp/sp & psp/sp | 1-21 | | | -0.5051 | -0.4447 | -0.4641 | -0.4571 | -0.3458 | | | |
| | | 2-32 | -0.3814 | -0.4675 | -0.3418 | | -0.3769 | | -0.3448 | -0.3850 | | |
| | | 3-43 | -0.4078 | -0.3627 | -0.3789 | -0.3566 | -0.3472 | -0.3910 | -0.4831 | -0.4364 | | |
| | fw/fr & fw/fr | 1-21 | -0.7557 | | -0.2499 | | | | -0.9223 | | -0.7099 | |
| | | 2-32 | -0.7228 | | -0.3218 | | -0.9787 | | -0.3036 | | -0.6554 | |
| | | 3-43 | -0.8160 | | | | -0.9489 | | -0.8892 | | -0.2862 | |
| | en/fr & fw/fr | 1-21 | | | | -0.2777 | | | | -0.3466 | -0.3281 | |
| | | 2-32 | | | -0.3616 | | | | | -0.2448 | | |
| | | 3-43 | | | -0.2887 | -0.3636 | | | | | -0.3117 | |
| | sp & fw/fr | 1-21 | | | | | | | | | 0.4282 | |
| | | 2-32 | | | | | | | | | 0.3371 | |
| | | 3-43 | | | | | | | | | | |
| | sp & sp | 1-21 | -0.4434 | -0.5196 | -0.3800 | -0.3362 | | | | | | -0.5632 |
| | | 2-32 | -0.3148 | -0.2960 | | | | | | | | -0.2833 |
| | | 3-43 | -0.2694 | -0.3216 | -0.5212 | -0.4178 | | | | | | -0.3647 |
| | en/fr & en/fr | 1-21 | | -0.3382 | | | | | | | | 0.2930 |
| | | 2-32 | | -0.2384 | | | | | | | | |
| | | 3-43 | | | | | | | | | | -0.2713 |
| | fw/fr & en/fr | 1-21 | | | 0.2576 | | | | | | | |
| | | 2-32 | | | | | | | | | | |
| | | 3-43 | | | 0.2632 | | | | | | | |

For the simplicity of the exposition, we explain the findings in each half of the table separately.

*4.2.1. The upper half*

In the upper half, we see the existence of an inverse linear relationship between en/fr and sp for men. This means the more increase in en/fr that men champions experience in one interval, the less tendency for self-presentation behavior they show in the next interval and vice versa. Moreover, by looking at the age analysis we see that this relationship gets bold at the age range-2, which is right at the middle of professional life for champions. The values in the second, third, fourth and fifth rows of the upper half show the existence of a relationship between the rate of change in one interval for a characteristic and the rate of change in the previous interval for the very same characteristic. Except for fw/fr the other three characteristics i.e. psp/sp, en/fr, and sp have an inverse relationship with their own previous rate of change for men champions. It means men have a balanced attitude toward these three characteristics to keep them away from monotonically increasing or decreasing over time, while for fw/fr, there is no tendency to balance it over time and this holds for all the age ranges. The other consistent behavior in the upper half is the inverse relationship between psp/sp and fw/fr and holds only for champions in age range 2 with no clue about their gender.

*4.2.2. The lower half*

In the first, second, fifth, and sixth rows of the lower half of the table, we see the existence of a relationship between the rate of change in one interval for a characteristic and the primary value of the same characteristic in the beginning of that interval. Here the inverse relationship manifests over all the four characteristics, and except en/fr whose inverse relationship only holds for men, the inverse relationship in the other three characteristics holds for both men and women. However, for sp this inverse relationship only is true for the older champions (age-range 3). The third row of the lower half of the table shows the existence of an inverse, linear and monotonic relationship between the rates of change in fw/fr in one interval with the primary value of en/fr in the beginning of the interval for women champions. It implies the fact that women champions who achieve higher levels of engagement from their followers would need less to follow others and vice versa, and this attitude gets bold by aging. Add to this, the last row of the lower half, ushers women champions who have higher values of fw/fr with an increase in en/fr in the upcoming interval.

## *4.3. Findings in brief*

Five main findings could be gleaned from the results, which are presented as follows.

1- The older the champions get, the less self-presenting photos (sp) they would post on Instagram. However, while men show an increase in their pure self-presenting posts (psp/sp), women become more reluctant to stand alone and take pure self-presenting posts.
2- As the champions get older, en/fr begins to decrease, and making more reciprocal ties with followers (increasing fw/fr) to draw more engagement from them (increasing en/fr) seems not to work well for men. Furthermore, while men experience a contraction in their fw/fr range, as they get older, women stretch it to the higher values. Indeed, aging makes the women champions to strengthen their links to the followers by reciprocation or even laying out new connections to non-followers.

3- The more increase in en/fr men champions experience in one interval, the less tendency for self-presentation behavior they show in the next interval and vice versa, and this relationship gets bold at the age range-2.

4- Except for fw/fr the other three characteristics i.e. psp/sp, en/fr, and sp have an inverse relationship with their own previous rate of change for men champions. It means men have a balance attitude toward these three characteristic to keep them away from monotonically increasing or decreasing over time, while for fw/fr not only men but women also show no tendency to balance it over time and this holds for all the age ranges. However, when it comes to the relationship between the rate of change in one interval for a characteristic and the primary value of the same characteristic in the beginning of the interval, the inverse relationship manifests over all the all four characteristics. However, except en/fr whose inverse relationship only holds for men, the inverse relationship in the other three characteristics holds for both men and women. Nonetheless, for sp this inverse relationship only is true for the older champions (age-range 3).

5- Female champions who achieve higher levels of engagement from their followers would need less to follow others and vice versa, and this attitude gets bold by aging.

## 5. Discussion

The results from this study offer several areas of discussion and implications that can be utilized by athletes and their sponsors as well as sport researchers. In the following, we will try to discuss each of the main findings (F1-5) in the light of related theories and previous findings of the research context.

## 5.1. *Impact of age and gender (with respect to RQ1)*

The finding about reduction of self-presentation by aging (F1) is quite in line with previous findings. Many researchers put self-presentation in the box of narcissism and came to this conclusion that narcissism decreases with age, since the narcissistic characteristic of not making commitments to others goes against the normative pathways (Roberts et al., 2010). Another study on NPI narcissism found a steady decrease in narcissism between age 15 and 54, with a small increase after age 55 (Foster et al., 2009). However, the increase in pure self-presenting posts for men champions relative to women champions as they age (F1) could be explained by their gender differences. Pure self-presentation means highlighting physical appearance and men are less concerned about their physical appearance (Gillespie and Eisler, 1992) and report higher appearance self-esteem (Gentile et al., 2009) than women. Therefore, women's lower confidence in their physical appearance leaves less tendency for them to stand alone in their presentation when they age.

The reason that champions receive less engagement and attention from their followers (F2) as they age hinges on the fact that the youth is the fuel of ambition and rivalry, which in turn induces users of an extended range to engage in the posts of their role model. Nevertheless, the question is why aging leads women champion to increase fw/fr, which stands in sharp contrast with what men champions do (F2). Here, the theory of evolutionary psychology sheds light into this gender difference. The core idea in evolutionary theory is natural selection, which states traits that support survival and reproduction have a higher chance to be passed to next generations (Gazzaniga et al., 2009). To make sure survival and reproduction are fulfilled, the Mother Nature has given different roles to men and women by which they define themselves. Women, as the main caregivers, develop skills for attachment, sensitivity and communication. Men, on the other hand, had to display power and dominance in order to maintain their status (Dovidio et al., 1988). Hence, the rise of fw/fr for women champions as they age is an attempt to keep the circle of communication strengthened when the power of youth would not come to help anymore.

## *5.2. The predictability of change (with respect to RQ2)*

We found that by increasing in en/fr in one interval the upcoming one bears a decrease in self-presentation for men champions and a reduction in fw/fr for women champions (F3 and F5). Here, the theory of uses and gratification (UGT) hand in hand with the theory of evolutionary psychology cast the spotlight on the reason behind these phenomena. The UGT explores how individuals deliberately seek out media to fulfill certain needs or goals (Katz et al., 1973), and gratification of needs is the most important role of media for humans. This leads users to choose a social media for a certain need and to satisfy that need they put their efforts into it, so it makes sense to see a moderation in the intensity of efforts when they reach to it. The higher rates of en/fr which implies the higher amount of attention from the followers would be of the major goals users seek for in social media and when is reached the moderation in the number of self-presenting photos in men champions' posts, and the reduction in the number of users the women champions follow, would occur. However, the question is why men try to craft their virtual self through self-presentation, while women try to do it with communication. Indeed, based on the theory of evolutionary psychology these gender differences are in large part socially determined over time as a result of the different reinforcements that men and women received for using particular self-presentation strategies. Men used to get admiration when they stood out with showing more privileges they had over their rivals, while women felt safety through the communication and establishing the channels of intimacy.

The up and down behavior of the characteristics in the consecutive time intervals for both men and women champions (F4) denotes the existence of an equilibrium between the amount of efforts users put into social media and the level of expectation they have to get their needs satisfied from the use of that social media. However, fw/fr seems not to fit in this equation as we see consecutive intervals of ascending or descending for the value of this characteristic.

Aside from contributions to the better understanding of virtual self for Olympic gold medalists in terms of Instagram characteristics and their interrelations, the findings of this research, may help athletes wishing to raise their public profile and build their personal brand by pursuing those strategies, which boost en/fr. Furthermore, as we can see from the literature (Casalo et al., 2017) the antecedents of users represent a lot about their behaviors, and findings of our research about user characteristics can pave the way to dig more in this line of research that incorporates cyber behaviors particularly for celebrities who have influence on their followers. Indeed, user categorization based on the characteristics in social media has been of great importance (Bodaghi and Oliveira, 2020; Tuna et al. 2016) and had a variety of applications as well in modelling of rumors and information spreading in online social networks (Liu et al., 2019; Bodaghi and Goliaei, 2018; Bodaghi et al., 2019).

## 6. Conclusion

By doing a longitudinal analysis on Instagram activities of Olympic gold medalists, we found an early understanding of their cyber behavioral diversity. In fact, the user characteristics open a window for us, through which we can study human behavioral features more precisely. The findings serve as a guide for sport researchers seeking to better understand athletes' use of social media, to interact with their followers and build the personal brand, which involves sponsorship and other business/promotional opportunities.

### *6.1. Limitations and future works*

Making sure an athlete is the owner of a given Instagram account was challenging when that account was not officially registered. It happened in a few cases, and we studied all their posts to make sure those accounts belong to them and for this we set the basis on the subject of photo posts, because it is highly likely that when one adult was shown in the photo, the photo belonged to that user (Tifferet and Vilnai-Yavetz, 2014). Due to the complexity of analysis in video posts, in this study, we only considered photo posts, however, an extension of this study may include video posts particularly by the help of AI, since it can be time-consuming for human analysis, and get deeper insights into the athletes' behavior on social media. Moreover, in this study, we showed how champions of different countries differ in terms of getting high/low ranks with respect to different characteristics, but the interpretation of these disparities in a standardized framework demands deep studies on the impact of cultures and nationalities on the use of social media, which in turn stimulates new avenues of research.

**Funding**

This research did not receive any specific grant from funding agencies in the public, commercial, or not-for-profit sectors.

## Disclosure statement

No potential conflict of interest was reported by the authors.